\documentclass[10.5pt,compsoc]{tst}

\usepackage{graphicx}
\usepackage{footmisc}
\usepackage{subfigure}
\usepackage{url}
\usepackage{multirow}
\usepackage[noadjust]{cite}
\usepackage{amsmath,amsthm}
\usepackage{amssymb,amsfonts}
\usepackage{booktabs}
\usepackage{color}
\usepackage{ccaption}
\usepackage{booktabs}
\usepackage{float}
\usepackage{fancyhdr}
\usepackage{caption}
\usepackage{xcolor,stfloats}
\usepackage{comment}
\graphicspath{{figures/}}
\usepackage{cuted}
\usepackage{captionhack}
\usepackage{epstopdf}
\usepackage{mathtools}
\usepackage{pifont}
\usepackage{xcolor}
\usepackage{tabularx}
\usepackage[numbers]{natbib}
\usepackage{hyperref}
\hypersetup{
  colorlinks = true, 
  citecolor = black, 
  linkcolor = black, 
  urlcolor = black
}

\headevenname{\normalsize{\textbf{\emph{Tsinghua Science and Technology, December}} 2022, 27(6): 881--893}}%
\headoddname{{\sf Jiashu Wu et al.:}\quad {\textbf{\emph{MIX-RS: A Multi-indexing System based on HDFS for Remote Sensing Data Storage}}}}%

\newtheoremstyle{mystyle}{0pt}{0pt}{\normalfont}{1em}{\bf}{}{1em}{}
\theoremstyle{mystyle}
\renewcommand\figurename{Fig.~}

\newcommand{\upcite}[1]{\textsuperscript{\cite{#1}}}

\newcommand{\nop}[1]{}

\newcommand*\numbercircled[1]{\raisebox{.5pt}{\textcircled{\raisebox{-.9pt} {#1}}}}

\newcommand{\modified}[1]{\textcolor{black}{#1}}

\makeatletter
\renewcommand{\@biblabel}[1]{[#1]\hfill}
\makeatother

\begin{document}

\hyphenpenalty=50000

\makeatletter
\newcommand\mysmall{\@setfontsize\mysmall{7}{9.5}}

\newenvironment{tablehere}
  {\def\@captype{table}}
  {}
\newenvironment{figurehere}
  {\def\@captype{figure}}
  {}

\thispagestyle{plain}%
\thispagestyle{empty}%

\let\temp\footnote
\renewcommand \footnote[1]{\temp{\normalsize #1}}
{}
\vspace*{-40pt}
\noindent{\normalsize\textbf{\scalebox{0.885}[1.0]{\makebox[5.9cm][s]
{TSINGHUA\, SCIENCE\, AND\, TECHNOLOGY}}}}

\vskip .2mm
{\normalsize
\textbf{
\hspace{-5mm}
\scalebox{1}[1.0]{\makebox[5.6cm][s]{%
I\hfill S\hfill S\hfill N\hfill{\color{white}%
l\hfill l\hfill}1\hfill0\hfill0\hfill7\hfill-\hfill0\hfill2\hfill1\hfill4
\hfill \color{black}{\quad 0\hfill 1\hfill /\hfill 1\hfill 0\quad p\hfill p\hfill  8\hfill 8\hfill 1\hfill --\hfill 8\hfill 9\hfill 3}\hfill}}}}

\vskip .2mm
{\normalsize
\textbf{
\hspace{-5mm}
\scalebox{1}[1.0]{\makebox[5.6cm][s]{%
DOI:~\hfill~\hfill1\hfill0\hfill.\hfill2\hfill6\hfill5\hfill9\hfill9\hfill/\hfill T\hfill S\hfill T\hfill.\hfill2\hfill0\hfill 2\hfill 1\hfill.\hfill9\hfill0\hfill1\hfill0\hfill 0\hfill 8\hfill 2}}}}

\vskip .2mm\noindent
{\normalsize\textbf{\scalebox{1}[1.0]{\makebox[5.6cm][s]{%
\color{black}{V\hfill o\hfill l\hfill u\hfill m\hfill%
e\hspace{0.356em}27,\hspace{0.356em}N\hfill u\hfill%
m\hfill b\hfill e\hfill r\hspace{0.356em}6,\hspace{0.356em}%
D\hfill e\hfill c\hfill e\hfill m\hfill%
b\hfill e\hfill r\hfill \hspace{0.356em}2\hfill0\hfill 2\hfill 2}}}}}\\

\begin{strip}
{\center
{\LARGE\textbf{MIX-RS: A Multi-indexing System based on HDFS for\\Remote Sensing Data Storage}}
\vskip 9mm}

{\center {\sf \large
Jiashu Wu, Jingpan Xiong, Hao Dai, Yang Wang$^*$ and Chengzhong Xu}
\vskip 5mm}

\centering{
\begin{tabular}{p{160mm}}

{\normalsize
\linespread{1.6667} %
\noindent
\bf{Abstract:} {\sf
A large volume of remote sensing (RS) data has been generated with the deployment of satellite technologies. The data facilitates research in ecological monitoring, land management and desertification, etc. The characteristics of RS data (e.g., enormous volume, large single-file size and demanding requirement of fault tolerance) make the Hadoop Distributed File System (HDFS) an ideal choice for RS data storage as it is efficient, scalable and equipped with a data replication mechanism for failure resilience. To use RS data, one of the most important techniques is geospatial indexing. However, the large data volume makes it time-consuming to efficiently construct and leverage. Considering that most modern geospatial data centres are equipped with HDFS-based big data processing infrastructures, deploying multiple geospatial indices becomes natural to optimise the efficacy. Moreover, because of the reliability introduced by high-quality hardware and the infrequently modified property of the RS data, the use of multi-indexing will not cause large overhead. Therefore, we design a framework called Multi-IndeXing-RS (\emph{MIX-RS}) that unifies the multi-indexing mechanism on top of the HDFS with data replication enabled for both fault tolerance and geospatial indexing efficiency. Given \modified{the fault tolerance provided by the HDFS, RS data is structurally stored inside for faster geospatial indexing.} Additionally, multi-indexing enhances efficiency. \modified{The proposed technique naturally sits on top of the HDFS to form a holistic framework without incurring severe overhead or sophisticated system implementation efforts.} The \emph{MIX-RS} framework is implemented and evaluated using real remote sensing data provided by the Chinese Academy of Sciences, demonstrating excellent geospatial indexing performance. }
}

\end{tabular}
}
\vskip 6mm

\vskip -3mm
\small\end{strip}

\thispagestyle{plain}%
\thispagestyle{empty}%
\makeatother
\pagestyle{tstheadings}

\begin{figure}[b]
\vskip -6mm
\begin{tabular}{p{44mm}}
\toprule\\
\end{tabular}
\vskip -4.5mm
\noindent
\setlength{\tabcolsep}{1pt}
\begin{tabular}{p{1.5mm}p{79.5mm}}
$\bullet$& Jiashu Wu, Jingpan Xiong and Hao Dai are with Shenzhen Institute of Advanced Technology, Chinese Academy of Sciences, Shenzhen 518055, China, as well as the University of Chinese Academy of Sciences, Beijing 100049, China. E-mail: \{js.wu, jp.xiong, hao.dai\}@siat.ac.cn
\\
$\bullet$& Yang Wang is with Guangdong-HongKong-Macao Joint Laboratory of Human-Machine Intelligence-Synergy Systems, Shenzhen Institute of Advanced Technology, Chinese Academy of Sciences, Shenzhen 518055, China. E-mail: yang.wang1@siat.ac.cn
\\
$\bullet$& Chengzhong Xu is with University of Macau, Macau 999078, China, China. E-mail: czxu@um.edu.mo
\\
$\sf{*}$&
To whom correspondence should be addressed. \\
          &          Manuscript received: year-month-day;
          accepted: year-month-day

\end{tabular}
\end{figure}\large


\section{Introduction}
\label{sec:sec1_introduction}

Advanced satellites from several space agencies\upcite{nasa_satellite,esa_satellite,landsat8_satellite} are constantly orbiting the planet, generating a massive amount of remote sensing (RS) data\upcite{huge_amount_rs_data_wang2020scalable,huge_amount_rs_data_huang2018agricultural,huge_amount_rs_data_huang_a,huge_amount_rs_data_huang_b}. The size of RS images is typically large as they are captured by top-tier camera devices with multiple bands or layers. As the RS data is produced daily\upcite{rs_data_daily_basis_liang2020estimating} in the era of big data\upcite{chen2014big_big_data_era,li2021self_big_data_era}, the order of magnitude of the data grows to terabyte\upcite{rs_data_tb_haut2021distributed} or even petabyte\upcite{rs_data_pb_warren2015seeing}. As a result, the Hadoop Distributed File System (HDFS) is often utilised as storage for the RS data\upcite{hdfs_for_rs_data_li2019improved,hdfs_for_rs_data_semlali2021big,hdfs_for_rs_data_xing2019intelligent,hdfs_wang2013rapid}. Deploying HDFS as the RS data storage can not only bring efficiency and scalability as the amount of data increases but also provide fault tolerance because of the data replication equipped by the HDFS\upcite{hdfs_data_replication_fault_tolerance_karun2013review}. 

However, the HDFS does not address the inefficiency of the geospatial indexing caused by the ever-growing data volume. Constant research efforts have been directed to optimise the efficacy of geospatial indexing and the utilisation of RS data\upcite{spatialhadoop_eldawy2015spatialhadoop,optimise_eff_spatial_idx_eldawy2013cg_hadoop,gisqf_al2014gisqf,shahed_eldawy2015shahed}. However, existing frameworks are inefficient because they used Hadoop MapReduce\upcite{spatialhadoop_eldawy2015spatialhadoop,gisqf_al2014gisqf,shahed_eldawy2015shahed}, which is time-consuming at start-up\upcite{mapreduce_overhead_ding2011more}, requires sophisticated effort to implement and change and does not attempt to use multi-indexing to optimise the performance. 

Given that the RS data is stored on HDFS with data replication enabled, we consider it natural to deploy a multi-indexing mechanism on top of the HDFS to form a unified framework for better geospatial indexing efficiency. The rationale of using the multi-indexing mechanism are as follows: 

\begin{itemize}

  \item RS data is stored on HDFS with data replication enabled, therefore, unifying the multi-indexing mechanism on top of the data replication is natural and feasible. The storage nodes can also parallelise the index construction and querying, boosting the indexing performance. 
  
  \item Using multiple light-weighted geospatial indexing algorithms makes the multi-indexing mechanism more resilient to single-point failures. Moreover, as different indexing algorithms may present different performances when processing queries that involve varying amounts of data, utilising the fastest one in the multi-indexing mechanism can boost the indexing performance when tackling different queries. 

  \item RS data is stored in modern data centres equipped with top-tier hardware infrastructures that are highly reliable and less frequent to fail\upcite{excellent_infra_lu2011review}. Therefore, geospatial indices will not be frequently re-constructed due to frequent hardware failures and hence utilising multiple geospatial indices will not cause severe overhead. 

  \item RS data also has characteristics of not being frequently modified, i.e., nearly read-only and is stored in a well-structured manner that benefits geospatial index construction and avoids constant index re-construction. As such, building multiple geospatial indices using relatively light-weighted indexing algorithms is feasible and suitable for RS data and will not cause severe time or space overhead. 

\end{itemize}

\modified{In this paper, considering the benefit and suitability of using the multi-indexing mechanism to improve geospatial indexing efficiency, we design a framework named \underline{M}ulti-\underline{I}nde\underline{X}ing - \underline{R}emote \underline{S}ensing (\emph{MIX-RS}).} For the multi-indexing mechanism, two popular and broadly-used geospatial indexing methods, i.e., \emph{GeoHash}\upcite{geohash_fox2013spatio,geohash_suwardi2015geohash,widely_used_geohash_huang2018rapid,widely_used_geohash_liu2014geohash}, \emph{QuadTree}\upcite{quadtree_whitman2014spatial,widely_used_quadtree_xu2020scienceearth} and the traditional indexing method \emph{Orthogonal List} are constructed and unified to form a multi-indexing ensemble. These indexing methods are lightweight, simple and less computation-intensive compared with other geospatial indexing methods, and hence will not compromise the efficiency in terms of time and space\upcite{widely_used_geohash_huang2018rapid,geohash_simple_petrov2018geohash,geohash_simple_quadtree_simple_guo2019geographic}. The \emph{MIX-RS} then unifies the multi-indexing mechanism on top of the HDFS with data replication. It can improve the geospatial indexing performance and benefit the applicability of RS data while not causing severe overhead. Moreover, the \emph{MIX-RS} framework does not require fundamental changes and only needs subtle implementation efforts, making it applicable to other applications\upcite{land_cover_change_mithal2013change,ocean_dynamics_faghmous2013parameter,climate_informatics_yu2013computational,application_traffic_forecasting,application_visualisation,dl_appl}. The prototype of the \emph{MIX-RS} framework is implemented, and is evaluated using real RS data provided by the Chinese Academy of Sciences\upcite{rceeca_cas,landsat8_satellite}, demonstrating superior indexing performance over compared indexing methods and frameworks. 

\modified{In summary, we make the following contributions in this paper: }

\begin{itemize}

  \item \modified{We design the \emph{MIX-RS} framework that naturally unifies the multi-indexing mechanism on top of the HDFS with data replication enabled, which aims for both fault tolerance and geospatial indexing efficiency improvement. }

  \item \modified{The proposed \emph{MIX-RS} framework improves the geospatial indexing performance and the applicability of RS data while not causing severe overhead or requiring sophisticated system implementation. }
  
  \item \modified{We implement the \emph{MIX-RS} framework and evaluate it using real RS data to validate its excellent geospatial indexing performance. }
  
\end{itemize}

The rest of this paper is organised as follows: Related geospatial indexing methods, as well as some frameworks that are used to index and query geospatial data, are introduced in Section \ref{sec:sec2_relatedwork}. By analysing these related methods, the rationale and motivation of the proposed \emph{MIX-RS} framework are introduced. The \emph{MIX-RS} framework is explained in detail in Section \ref{sec:sec3_the_proposed_framework}. Section \ref{sec:sec4_experiments_and_results_analysis} presents the empirical evaluation and performance analysis of the \emph{MIX-RS} framework. Section \ref{sec:sec5_conclusion} concludes the paper.


\section{Related Work and Opportunity}
\label{sec:sec2_relatedwork}

As the scale of RS data keeps growing rapidly, how to efficiently index the data to satisfy user queries becomes a crucial problem, and thereby has attracted attention from both industry and academic communities. In this section, we first overview widely used geospatial indexing algorithms, then present proposed frameworks that deal with efficient geospatial storage and indexing. Finally, we identify their deficiencies to show the motivations and research opportunities of our proposed multi-indexing mechanism and the \emph{MIX-RS} framework. 

\subsection{Geospatial Indexing Algorithm}
\label{sec:sec2.1_geo_spatial_indexing_algorithms}

Aji et al.,\upcite{aji_aji2013hadoop} and Eldawy et al.,\upcite{spatialhadoop_eldawy2015spatialhadoop} proposed the Uniform Grid Index, which was one of the most commonly used geospatial indexing algorithms. The Uniform Grid Index performed geospatial indexing by constructing an index table based on the longitude and latitude coordinates. However, the algorithm needed to traverse the entire index table when looking for a specific coordinate, resulting in tremendous space and time consumption. If, in addition, temporal information is added as a new indexing dimension, then the storage complexity will be severely raised, and the searching and indexing efficiency will be heavily impaired. 

\begin{figure}[!h]
  \begin{center}
    \includegraphics[width=0.45\textwidth]{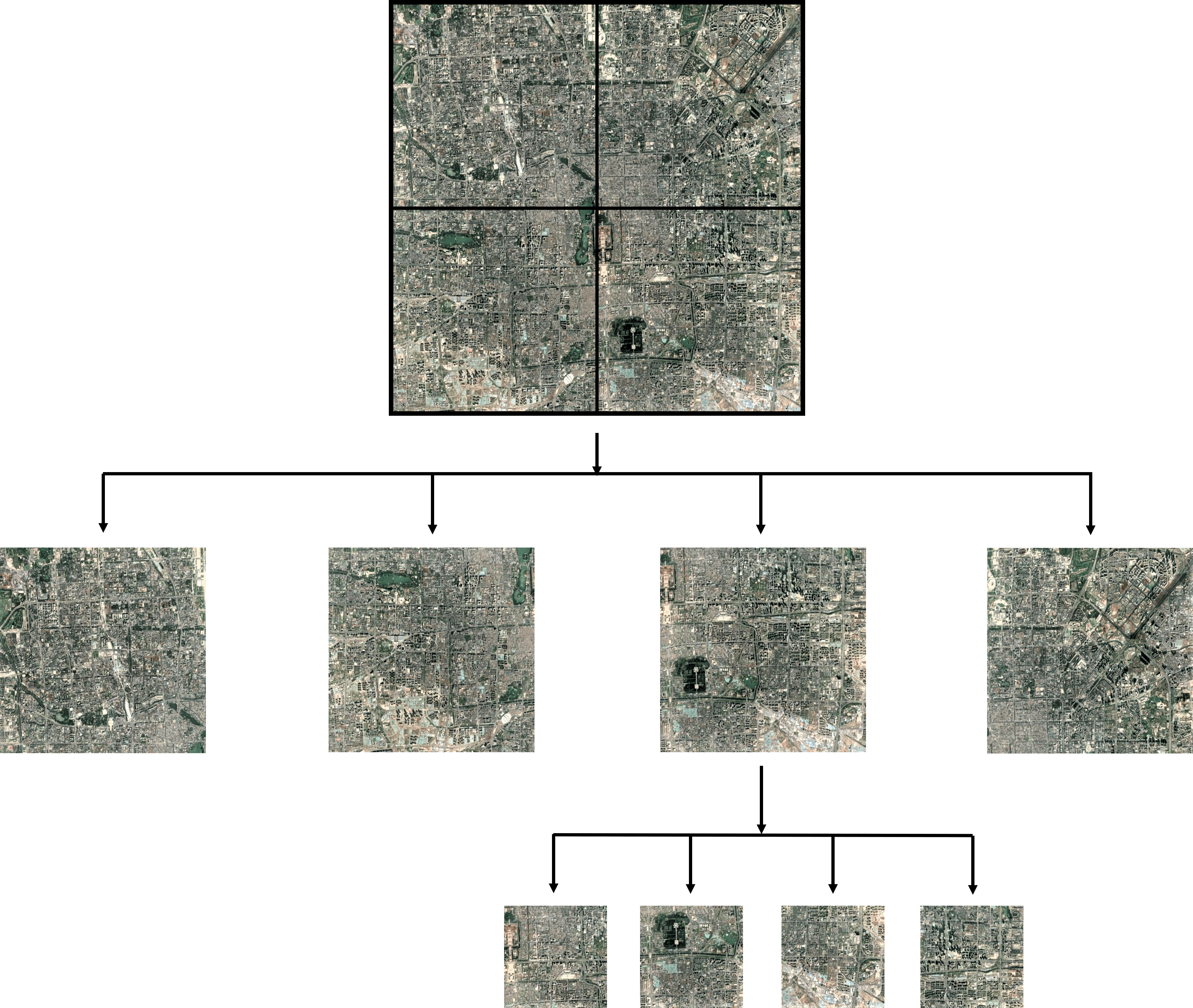}\\
    \caption{\modified{An illustrative example of the Quadtree algorithm when storing the RS data of City of Beijing, the capital of China. The RS data is divided into four quadrants and stored into a tree-like structure, where each quadrant represents a sub-region of the city. Each quadrant is then further divided in the same way, forming a QuadTree with several layers. }}
    \label{fig:quadtree}
  \end{center}
\end{figure}

\modified{Whitman et al.,\upcite{quadtree_whitman2014spatial} put forward the QuadTree indexing algorithm. As illustrated in Fig. \ref{fig:quadtree}, the RS data is commonly divided into four quadrants by major RS applications such as Google Earth, which makes the QuadTree indexing algorithm very suitable to index RS data. The QuadTree indexing algorithm will transform the quadrants into a tree-like structure, and each quadrant is further divided in the same way, forming a QuadTree with several layers, as shown in Fig. \ref{fig:quadtree}. Since each sub-tree in the QuadTree indexing structure contained a piece of the sub-region of the globe, QuadTree was very efficient when performing regional queries. Compared with the Uniform Grid Index algorithm that simply traversed the longitude and latitude index, the QuadTree possessed better efficiency. Moreover, the space overhead caused by the QuadTree indexing algorithm was not heavy. }

\begin{figure}[!h]
  \begin{center}
    \includegraphics[width=0.45\textwidth]{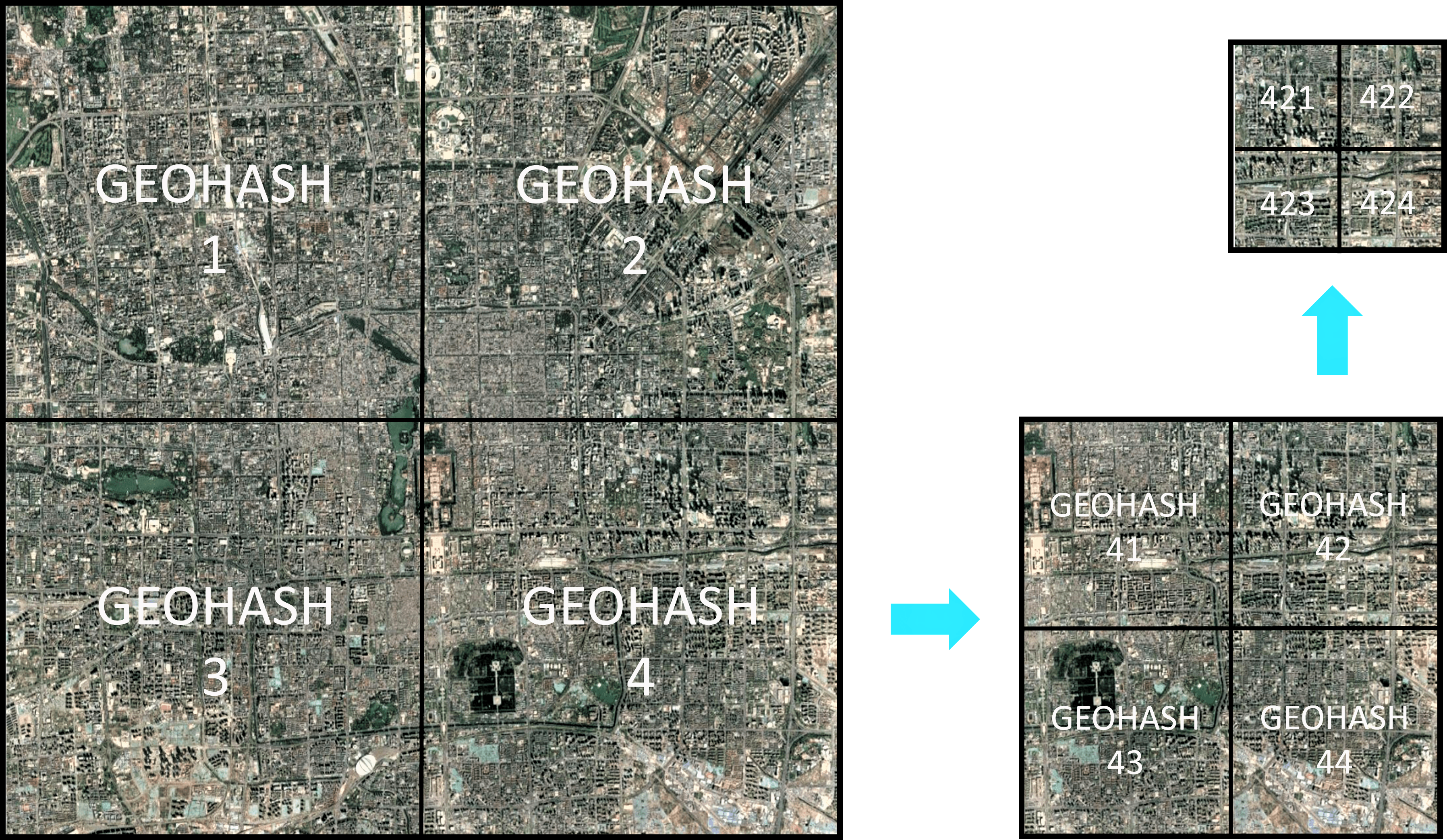}\\
    \caption{\modified{An illustrative example of the GeoHash algorithm when processing the RS data of City of Beijing. In the lower-right square, Forbidden City (with GeoHash ``41'') and Beijing Central Business District (CBD, with GeoHash ``42'') share the same GeoHash prefix ``4'' as they are spatially close to each other. In the smaller upper-left square, Forbidden City (with GeoHash ``411'') and Qianmen Business District (with GeoHash ``413'') share a longer common GeoHash prefix which is ``41'', since they are geographically closer to each other than the distance between the Forbidden City and Beijing CBD. Hence, GeoHash encodes a geographic location into a string so that the longer the shared prefix between two geohashes, the spatially closer they are together. The GeoHash codes in this example are for illustration purposes only. }}
    \label{fig:geohash}
  \end{center}
\end{figure}

\begin{figure*}[!h]
  \begin{center}
    \includegraphics[width=0.66\textwidth]{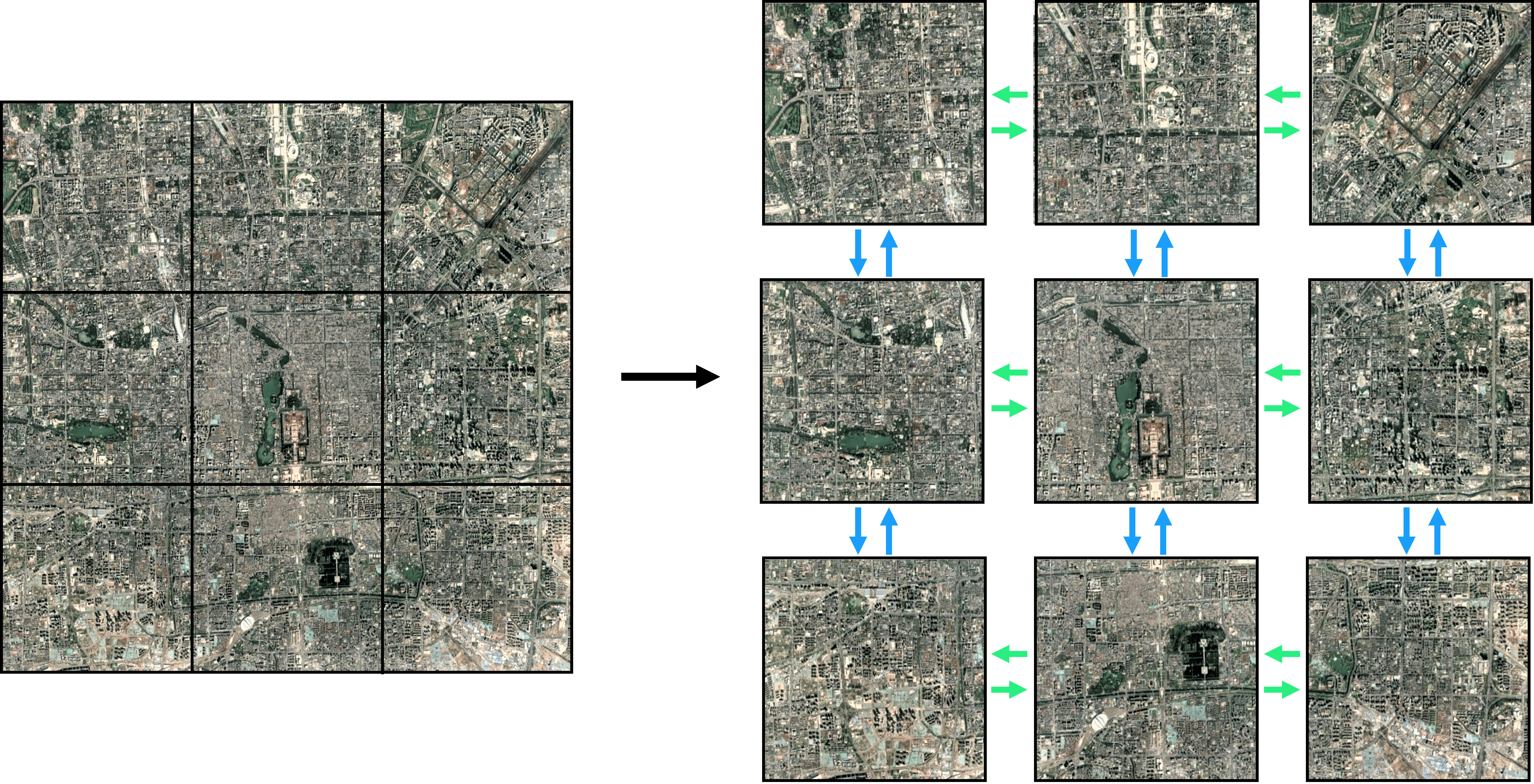}\\
    \caption{\modified{An illustrative example of the Orthogonal List algorithm when processing the RS data of City of Beijing. The algorithm will store the RS data into an orthogonal linked list, in which grids are arranged using their coordinate location. }}
    \label{fig:orthogonal_list}
  \end{center}
\end{figure*}

\modified{Fox et al.,\upcite{geohash_fox2013spatio} presented the GeoHash indexing algorithm. GeoHash encoded a geographic location into a string containing letters and digits so that the longer the shared prefix between two geohashes, the spatially closer they are together. The binary coordinate encoding was utilised to convert a coordinate range into a string of binary numbers, which was then grouped into several $6$-digit groups, and finally converted using Base$32$ encoding. An illustrative example of City of Beijing is presented in Fig. \ref{fig:geohash}. In this example, the Forbidden City (North-West in the lower-right square) and Beijing CBD (North-East in the lower-right square) will have their geohashes start with the common prefix ``4''. Inside the sub-region with geohash ``41'', the Forbidden City and the Qianmen Business District will share a common prefix ``41'' since they are spatially closer to each other. The GeoHash was highly efficient as it reduced the length of the encoding to be stored, leading to significantly better performance, especially when performing regional RS data indexing. }

\modified{Another widely used algorithm suitable to store data with orthogonal information is the Orthogonal List algorithm\upcite{orthgonal_list_introduction_to_algorithms_cormen2009introduction}. Although not specifically designed for geospatial indexing, the Orthogonal List algorithm can still be a suitable candidate for RS data storage and indexing as its structure can perfectly fit the orthogonal structure such as the longitude and latitude coordinate of a geographic location. As illustrated in Fig. \ref{fig:orthogonal_list}, each piece of RS data is linked to its geographically-adjacent region in the Orthogonal List data structure, which makes the Orthogonal List a natural choice to process RS data. }

\subsection{Geospatial Storage and Indexing Framework}
\label{sec:sec2.2_geo_spatial_indexing_frameworks}

Several research efforts leveraged the aforementioned popular geospatial indexing algorithms to design geospatial storage and indexing frameworks. Eldawy et al.,\upcite{spatialhadoop_eldawy2015spatialhadoop} proposed an RS data storage and indexing framework named SpatialHadoop, which utilised their own designed Uniform Grid Index algorithm. The SpatialHadoop used the HDFS cluster as the data storage system, and then built an index on the upper layer of its file system and constructed a MapReduce interface to serve external requests. However, SpatialHadoop suffered from drawbacks caused by executing the MapReduce program, which required a start-up time that downgraded the performance. Furthermore, SpatialHadoop required nearly 14k lines of code based on Hadoop, which made the implementation very labourious. 

The MapReduce-based SHAHED framework was also proposed by Eldawy et al.,\upcite{shahed_eldawy2015shahed} to query, visualise and mine large-scale RS data generated by the satellites. Unlike their previous work, SHAHED leveraged QuadTree as its indexing method. The SHAHED considered both spatial and temporal aspects of the RS data for effective querying. The query component of the SHAHED constructed the index for querying, and the visualisation component used the MapReduce program to generate the heatmap related to the user query. Despite these efforts, the SHAHED still suffered from the warm-up time overhead caused by the MapReduce program, which reduced its performance. 

Al-Naami et al.,\upcite{gisqf_al2014gisqf} introduced the Geographic Information System Querying Framework (GISQF) that worked on top of the SpatialHadoop framework. The GISQF used a two-layer geospatial indexing strategy to accelerate query processing. However, the two-layer geospatial indexing not only consumed a tremendous amount of construction time and storage resources, but also required a sophisticated implementation. 

\subsection{Motivation and Research Opportunity}
\label{sec:sec2.3_motivation_and_research_opportunity}

By analysing these aforementioned frameworks, the HDFS was utilised for data storage by all of them. However, these frameworks only used a single geospatial indexing method during index construction and querying, which limited the merit of HDFS parallelism and the performance could be limited by utilising only a single index. Moreover, to efficiently utilise MapReduce, these frameworks required sophisticated implementation efforts to make significant modifications to the MapReduce framework. Hence, it naturally leads to the idea of unifying multi-indexing mechanisms on top of the HDFS as it not only enjoys the benefit of parallelism possessed by the HDFS data replication to speed up the remote sensing data indexing process, but also requires subtle changes and saves the laborious implementation efforts. 

In terms of the proposed multi-indexing mechanism, we utilise two widely-used geospatial indexing algorithms, i.e., GeoHash and QuadTree, and a traditional data indexing algorithm, i.e., Orthogonal List. The reasons for adopting these indexing algorithms are as follows: 

\begin{itemize}

  \item The GeoHash and QuadTree algorithms are commonly used geospatial indexing algorithms for RS data, such as in\upcite{widely_used_geohash_huang2018rapid,widely_used_geohash_liu2014geohash,widely_used_quadtree_xu2020scienceearth}. Although the Orthogonal List is not specifically designed for geospatial indexing, its orthogonal linking structure is a suitable solution to store RS data that possesses orthogonal information such as longitude and latitude. 
  
  \item Unlike other indexing algorithms, such as Uniform Grid Index and R-tree\upcite{rtree_overhead_zhang2018efficient}, the GeoHash, QuadTree and Orthogonal List algorithms cause subtle time and space overhead, as they do not require complicated indexing data structures. These algorithms are lightweight, simple and not computation-intensive, which will not compromise efficiency. 
  
  \item These three indexing algorithms are relatively easy to implement and deploy, requiring lighter implementation efforts. 

\end{itemize}

Hence, by unifying the multi-indexing mechanism on top of the HDFS with data replication, the system efficiency can be increased while not causing severe time or space overhead, and meanwhile avoiding complicated implementation efforts.


\begin{figure*}[!t]
  \begin{center}
    \includegraphics[width=0.9\textwidth]{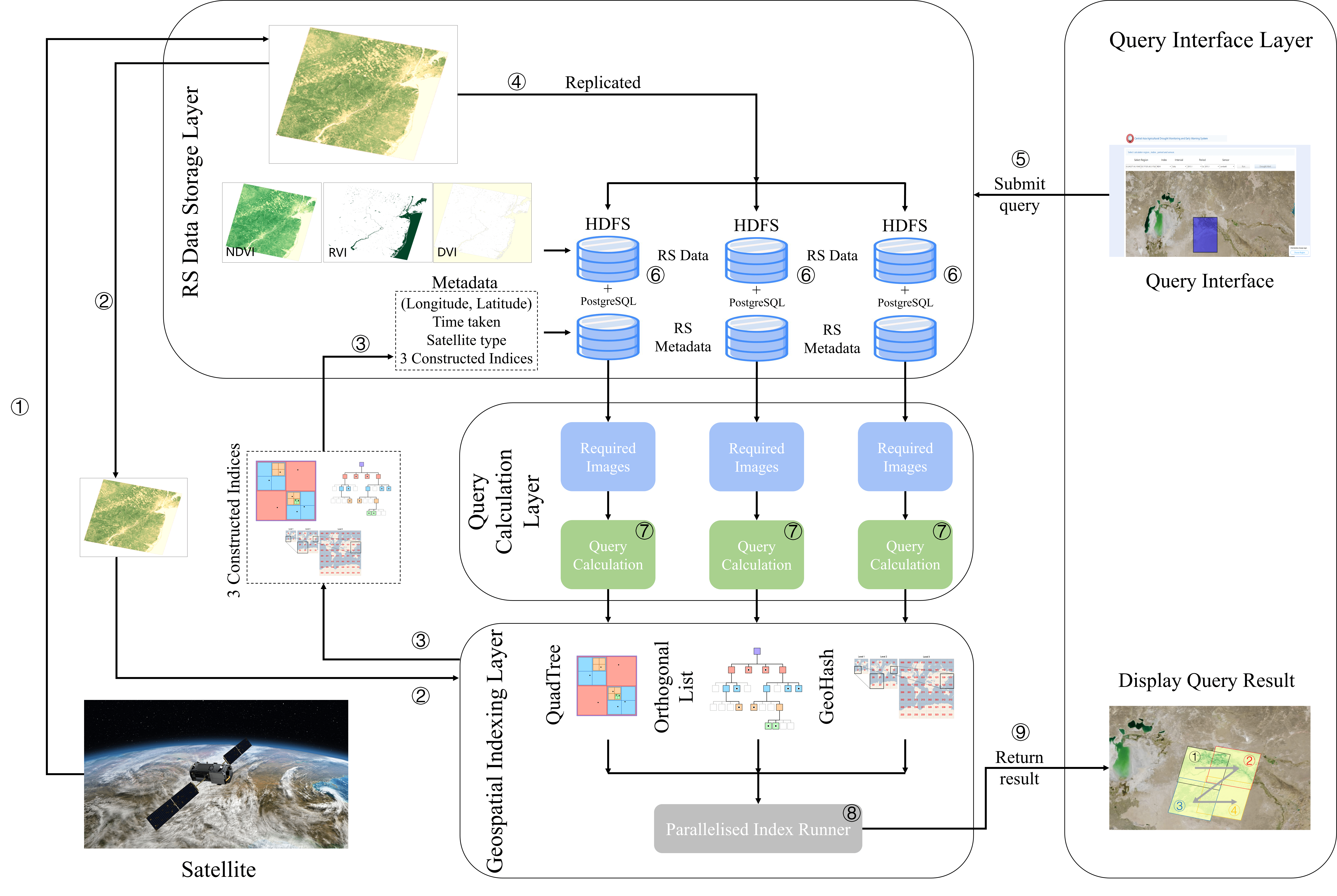}\\
    \caption{The \emph{MIX-RS} framework. The framework constitutes four layers, i.e., the Remote Sensing (RS) Data Storage Layer, Query Interface Layer, Query Calculation Layer and Geospatial Indexing Layer. Step \numbercircled{1}: the RS data is downloaded to the RS Data Storage Layer for preprocessing and storage; Step \numbercircled{2}: the RS data is directed to the Geospatial Indexing Layer to construct three geospatial indices; Step \numbercircled{3}: the three constructed geospatial indices are directed back to the RS Data Storage Layer and are treated as part of the metadata; Step \numbercircled{4}: the preprocessed RS data and its metadata are replicated. The RS data is stored into the HDFS structurally based on its geographical coordinate to form a tree structure, as this kind of directory structure can improve the data retrieval and indexing performed in later steps. The RS metadata is stored into the PostgreSQL database; Step \numbercircled{5}: the submitted user query will be sent from the Query Interface Layer to the RS Data Storage Layer; Step \numbercircled{6}: the PostgreSQL database will decide which pieces of RS data are involved in the range specified by the submitted query based on the RS metadata. Then, the required RS data is retrieved from the HDFS and used by the Query Calculation Layer; Step \numbercircled{7}: The Query Calculation Layer calculates the required information; Step \numbercircled{8}: three geospatial indexing algorithms will be run in parallel, and the result produced by the fastest one will be utilised; Step \numbercircled{9}: The result will be sent to the Query Interface Layer for visualisation. }
    \label{fig:figure1}
  \end{center}
\end{figure*}

\section{MIX-RS Framework and Workflow}\label{sec:sec3_the_proposed_framework}

\modified{In this section, we describe the proposed \emph{MIX-RS} approach in terms of its framework and workflow with a focus on how the multi-indexing is designed on top of the HDFS to improve the geospatial indexing efficiency. }

\subsection{MIX-RS Framework}
\label{sec:sec3.1_mix_rs_workflow}

The framework of the proposed \emph{MIX-RS} is illustrated in Fig. \ref{fig:figure1}. Its workflows between constituting layers are as follows: 

\emph{Remote Sensing Data Storage Layer}: The RS data captured by satellites is handed over to the RS Data Storage Layer, which serves as a preprocessing component to preprocess the data, including image-band separation and metadata extraction. The multi-index of the RS data are then constructed by triggering the Geospatial Indexing Layer, and the constructed multi-index are regarded as part of the metadata. Each piece of RS data with its metadata is replicated into three replicas. The RS data is stored in the underneath HDFS structurally using its geographical coordinate to form a tree directory structure so that it can benefit the data retrieval and indexing performed later. The RS metadata is stored in PostgreSQL database that will be used to decide involved RS data upon receiving the user query. (Section \ref{sec:sec3.2_remote_sensing_data_storage_layer})

\emph{Query Interface Layer}: Users interact with the Query Interface Layer and submit their query by specifying the region using longitude, latitude and time period. The users also need to specify the type of information that they want. The query is then sent to the RS Data Storage Layer for data retrieval, and the results produced by the Geospatial Indexing Layer are displayed to the users via the Query Interface Layer. An example query is illustrated in Fig. \ref{fig:figure2}. (Section \ref{sec:sec3.3_query_interface_layer})

\emph{Query Calculation Layer}: Once the data involved in the submitted query is retrieved by the RS Data Storage Layer, the Query Calculation Layer will process it to produce the required information like the vegetation rate in our experiment, or to calculate the information such as drought rate, etc. (Section \ref{sec:sec3.4_query_calculation_layer})

\emph{Geospatial Indexing Layer}: Upon receiving the resulting RS data from the Query Calculation Layer, three geospatial indexing algorithms are used to transform the RS images to form a holistic view of the area that they cover. For better efficiency, each indexing method will be conducted on one of the data replicas in the HDFS for parallelised computation, and the outcome produced by the fastest method will be utilised as the result. (Section \ref{sec:sec3.5_spatial_indexing_layer})

\subsection{MIX-RS Workflow}
A detailed explanation of each constituting layer in this section is provided using the framework design, i.e., the Remote Sensing Data Storage Layer, the Query Interface Layer, the Query Calculation Layer and the Geospatial Indexing Layer. 

\subsubsection{Remote Sensing Data Storage Layer}
\label{sec:sec3.2_remote_sensing_data_storage_layer}

The Remote Sensing Data Storage Layer in the \emph{MIX-RS} framework is used to preprocess, replicate and store the RS data received from the satellite. As shown in Step \numbercircled{1} in Fig. \ref{fig:figure1}, the RS data is downloaded from the satellite. Since the RS data composes of several bands such as Normalised Difference Vegetation Index(NDVI), Ratio Vegetation Index (RVI), Difference Vegetation Index (DVI) etc., hence during preprocessing, this $3$D RS data will be flattened into $2$D so that bands are separated. The metadata of the RS images including longitude, latitude, time taken and satellite type are also extracted. To prepare for the multi-indexing mechanism used later, the RS image metadata will be processed by the Geospatial Indexing Layer to construct three indices as shown in Step \numbercircled{2}. These three constructed indices are then sent back in Step \numbercircled{3} illustrated in Fig. \ref{fig:figure1} and are treated as part of the metadata. Finally, as indicated by Step \numbercircled{4}, the RS data and its metadata will then be replicated for the purpose of fault tolerance. The RS data will be stored into HDFS structurally using its geographical coordinate to form a tree directory structure, since such kind of structure can benefit the data retrieval and indexing performed later. The RS metadata is stored in the PostgreSQL database. 

\begin{figure*}[!t]
  \begin{center}
    \includegraphics[width=0.97\textwidth]{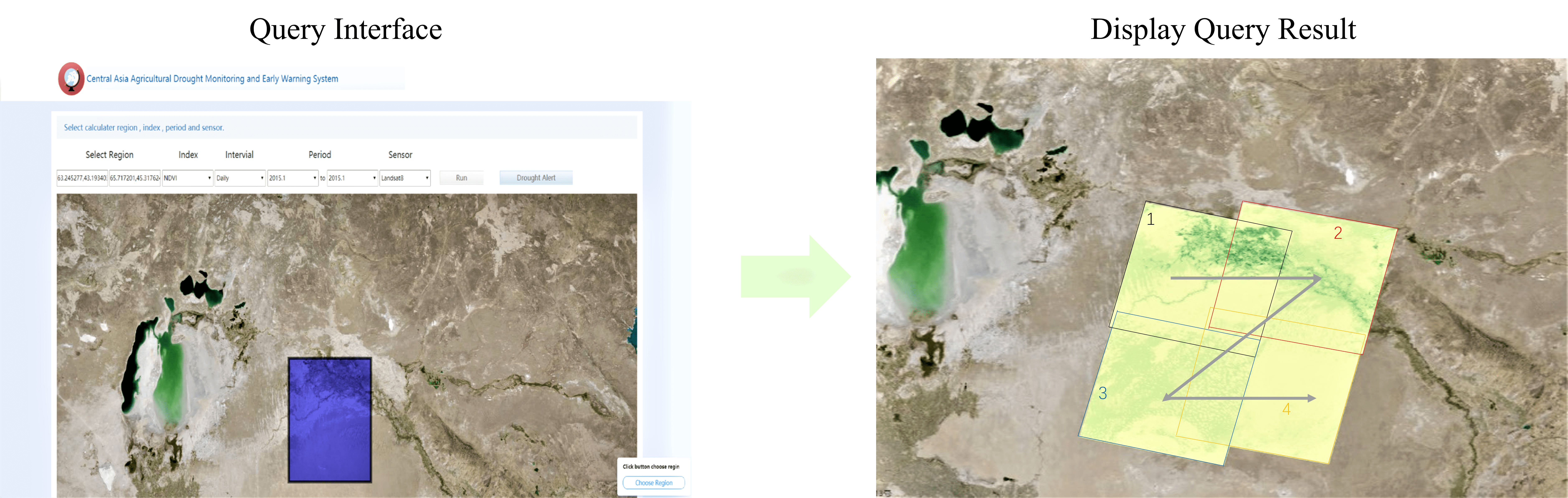}\\
    \caption{Illustration of an example query. The user queries the region of interest in the query interface on the left by specifying the longitude, latitude, time period and image source. The user also needs to specify the type of information they need, in this example, it is NDVI. The user query is then processed by the \emph{MIX-RS} framework. In this example, the specified region involves four RS images, hence the Remote Sensing Data Storage Layer needs to firstly retrieve the images involved in the query. The retrieved images are then processed by the Query Calculation Layer to calculate the required information. After calculating the information that the user needs, these four remote sensing images are placed in their designated position based on their geographic metadata by the multi-indexing mechanism. Finally, the constructed query result is returned and visualised by the user interface, as shown on the right-hand side. }
    \label{fig:figure2}
  \end{center}
\end{figure*}

Additionally, as indicated by Step \numbercircled{6} in Fig. \ref{fig:figure1}, the Remote Sensing Data Storage Layer is also responsible for retrieving the RS data required by the submitted query, i.e., RS data that covers the specified geographical range. For instance, in Fig. \ref{fig:figure2}, the specified blue user query shown on the left is covered by four pieces of RS data as shown on the right. The submitted query will specify the longitude and latitude range of interest, which is used by the PostgreSQL database to determine which pieces of RS data are required based on the metadata stored in it. Once the involved RS data is decided, it will be retrieved from the HDFS. 

\subsubsection{Query Interface Layer}
\label{sec:sec3.3_query_interface_layer}

Once the data preprocessing and storage are completed by the Remote Sensing Data Storage Layer (Steps \numbercircled{1} \& \numbercircled{4}) and the multi-index are constructed by the Geospatial Indexing Layer (Steps \numbercircled{2} \& \numbercircled{3}), the system is ready to receive queries via the Query Interface Layer. Fig. \ref{fig:figure2} presents the query interface and an example user query. When submitting the query, the user needs to specify the longitude and latitude of the area that they want to retrieve. Furthermore, the time period and type of information they are interested in will also be specified. The user query interface is illustrated on the left-hand side of Fig. \ref{fig:figure2}. The query will then be submitted to the Remote Sensing Data Storage Layer, as shown in Step \numbercircled{5} in Fig. \ref{fig:figure1}. In the end, the final result will be returned to the Query Interface Layer in Step \numbercircled{9} in Fig. \ref{fig:figure1} for result display and visualisation, as shown on the right-hand side of Fig. \ref{fig:figure2}. 

\subsubsection{Query Calculation Layer}
\label{sec:sec3.4_query_calculation_layer}

As illustrated in Fig. \ref{fig:figure1}, the Remote Sensing Data Storage Layer will prepare the required RS data for the Query Calculation Layer. The Query Calculation Layer can calculate different types of geospatial information (Step \numbercircled{7}), such as the NDVI as shown in Equation \ref{equ:equation1_ndvi} which is used in our evaluation. Note that NDVI is one of the indicator metrics that are valuable to assessing the rate of vegetation coverage to reflect the vegetation and nutrition condition of the area. The NIR is the infrared reflection rate and RED is the red-light reflection rate. Both the NIR and RED are two out of ten bands in every piece of RS data. The calculation is performed pixel-wise, and the result will still be a matrix with the same dimension as each original band of the RS data. The calculations are also performed in parallel between each processing node of the HDFS. 

\begin{equation}
  \label{equ:equation1_ndvi}
  NDVI = \frac{NIR - Red}{NIR + Red}
\end{equation}

\subsubsection{Geospatial Indexing Layer}
\label{sec:sec3.5_spatial_indexing_layer}

The idea of multi-indexing is applied in the Geospatial Indexing Layer to speed up the construction of the result. The calculated result produced by the Query Calculation Layer will be indexed by three geospatial indexing methods, i.e., GeoHash, QuadTree and Orthogonal List, and are performed in parallel. As shown in Step \numbercircled{8} in Fig. \ref{fig:figure1}, the parallelised index runner monitors the indexing progress of each processing node and utilises the fastest one that produces the result. By leveraging the multi-indexing as an ensemble on top of the HDFS with data replication, the delay on one or two of the processing nodes will not affect the overall performance since the fastest indexing will produce the result to satisfy the user query. Hence, it is reasonable that the multi-indexing mechanism can boost the overall system performance and outperform those systems that only use a single index, even though the data replication of the HDFS provides a perfect prerequisite to utilise multi-indexing. Once the result is generated, it will be returned to the Query Interface Layer in Step \numbercircled{9} as shown in Fig. \ref{fig:figure1}. 

\subsection{Prototype Implementation Detail}
\label{sec:sec_prototype_implementation_detail}

\begin{figure*}[!ht]
  \begin{center}
    \includegraphics[width=0.8\textwidth]{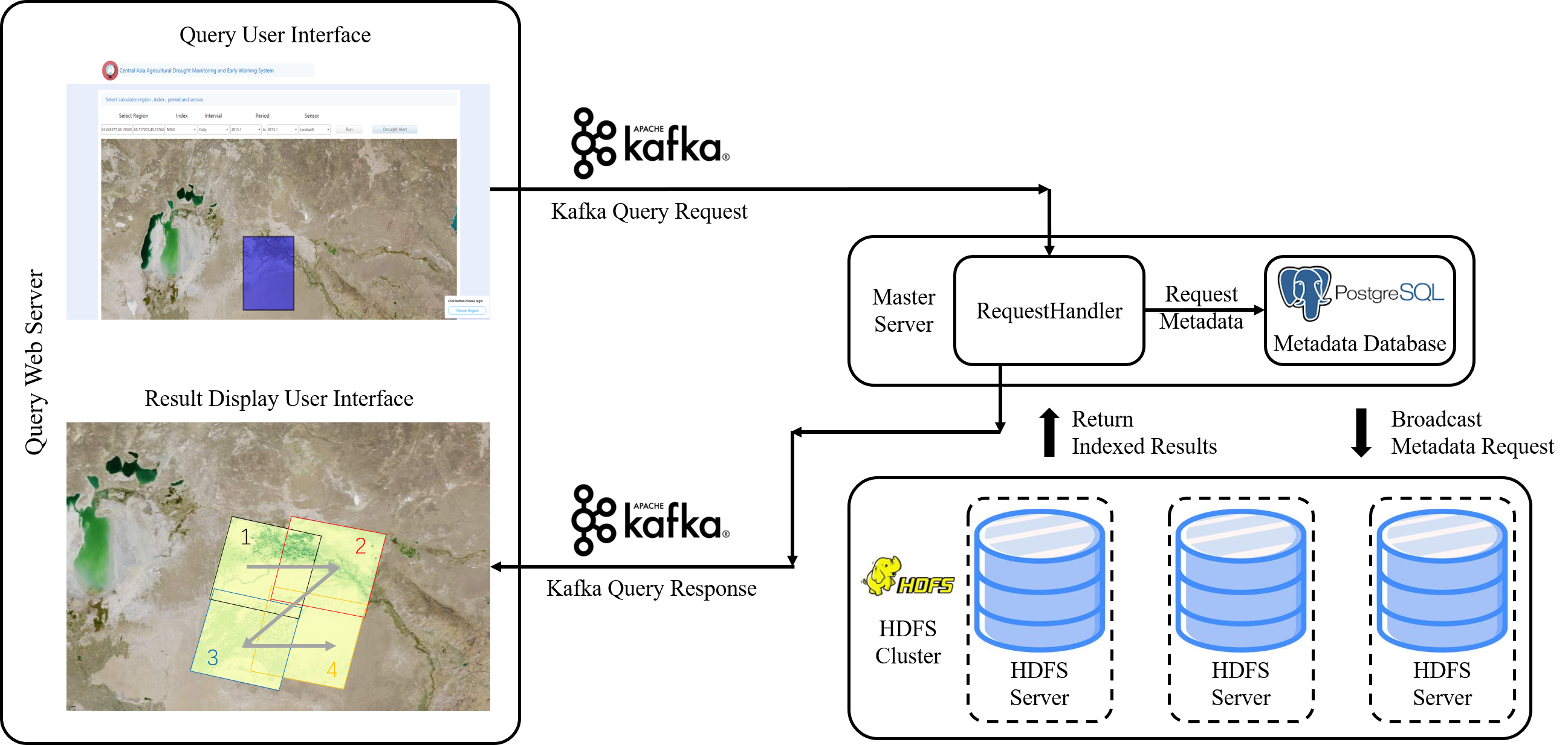}\\
    \caption{\modified{Illustration of the \emph{MIX-RS} prototype implementation. }}
    \label{fig:figure_prototype_implementation}
  \end{center}
\end{figure*}

\vspace{-0.1cm}

\modified{To verify the effectiveness of our proposed \emph{MIX-RS} framework and to make it become applicable in practice, we follow the architecture as depicted in Fig. \ref{fig:figure1} to implement the proposed \emph{MIX-RS} framework with corresponding designed layers and steps. The details of the prototype implementation are illustrated in Fig. \ref{fig:figure_prototype_implementation}. }

\modified{To receive the user queries, we implement a web-based user interface that allows the users to specify the region of interest by giving the longitude and latitude range of the region. To complete the query, the users also need to provide the time period, image source and the type of information they need. The submitted query is then sent to the RequestHandler residing in the master server via Kafka request message. Upon receiving the user query, the RequestHandler will ask the PostgreSQL metadata database to determine which pieces of RS data are involved in this query. Once completed, the master server broadcasts the information of the required data to the HDFS servers for data retrieval and geospatial indexing. Three HDFS servers will execute the geospatial indexing in parallel, and the fastest result received by the RequestHandler from these HDFS servers will be utilised and sent back to the user interface via Kafka messaging. Finally, the result will be displayed on the web user interface. }


\section{Empirical Evaluation}
\label{sec:sec4_experiments_and_results_analysis}

To validate the effectiveness of the \emph{MIX-RS} framework, comprehensive evaluations are performed on real RS data captured by the LandSat8 satellite. We verify the superiority of both the multi-indexing mechanism, which is compared with the single-indexing techniques, and the overall \emph{MIX-RS} system, which is compared with other widely-used systems such as SpatialHadoop\upcite{spatialhadoop_eldawy2015spatialhadoop}, GISQF \upcite{gisqf_al2014gisqf} and SHAHED\upcite{shahed_eldawy2015shahed}. 

\subsection{Experimental Setup and Dataset}
\label{sec:sec4.1_experimental_setup_and_dataset}

To verify the efficacy of the \emph{MIX-RS} system, real RS data captured by the LandSat8 satellite is utilised. The data is from the Central Asian Ecology and Environment Research Centre of the Chinese Academy of Sciences\upcite{rceeca_cas,landsat8_satellite} and contains about $9000$ pieces of RS data, with a total size of around 4TB. The RS data is in the \textit{geotiff} file format, with a dimension of $7000\times7000\times10$. \modified{Each experiment is repeated 50 times and the corresponding results have been plotted with error bars. }

\modified{In terms of system hardware configurations, the \emph{MIX-RS} system is deployed on five servers, one as the HDFS master node and three as the HDFS slave nodes, where each HDFS slave node stores one replica of the RS data. One extra server is used to deploy the query user interface. All servers used during experiments have the same hardware and OS configuration as shown in Table \ref{tab:table1}. In terms of software configurations, all servers have Kafka 2.3.1 installed to build a message queue for query-result communications, and have PostgreSQL 11.2 installed to serve as the RS metadata database. Besides, all HDFS servers have Hadoop 2.9.2 installed. }

\begin{table}[!t]
  \caption{Hardware and operating system (OS) configuration of the server infrastructure (The same configuration applies for all servers)}
  \centering
  \vspace*{0.3cm}
  \begin{tabularx}{0.45\textwidth}{lX}
  \toprule
  \bfseries Item & \bfseries Configuration\\
  \hline
  CPU & Intel(R) Xeon(R) CPU \newline E5-2630 v4@2.20GHz \newline (10 Cores, 20 Threads)\\
  Number of CPUs & 2\\
  Memory Size & 64GB\\
  Disk Space & 25TB\\
  OS & Ubuntu 16.04.5 LTS\\
  \hline
  Cluster Configuration & 1*HDFS Master node + \newline 3*HDFS Slave nodes + \newline 1*Query server\\
  \bottomrule
  \end{tabularx}
  \label{tab:table1}
\end{table}

\begin{figure*}[!t]
  \begin{center}
    \includegraphics[height=5.5cm,keepaspectratio]{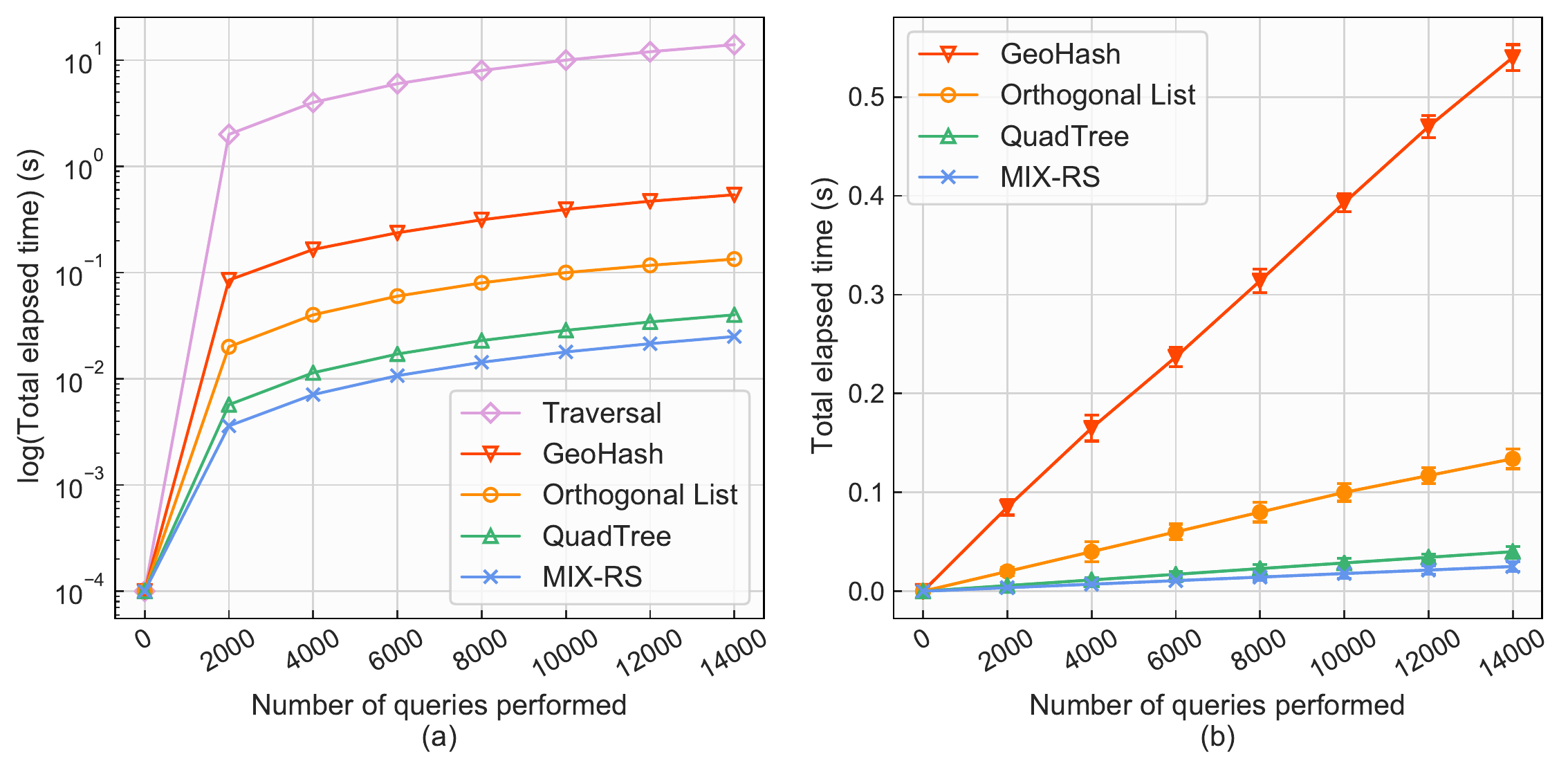}\\
    \caption{Performance comparisons between the multi-indexing and single indexing methods when processing different number of queries. The X-axis indicates the number of user queries, and the Y-axis indicates the total elapsed time. Since the performance of the brute force traversal method is much worse than other methods, hence in (a), log-scale is applied on the Y-axis. To better visualise the relationships between methods and the linear trends, the brute force traversal is eliminated in (b) and the log-scale in Y-axis is also removed. }
    \label{fig:figure3}
  \end{center}
\end{figure*}

\begin{figure*}[!t]
  \begin{center}
    \includegraphics[height=5.5cm,keepaspectratio]{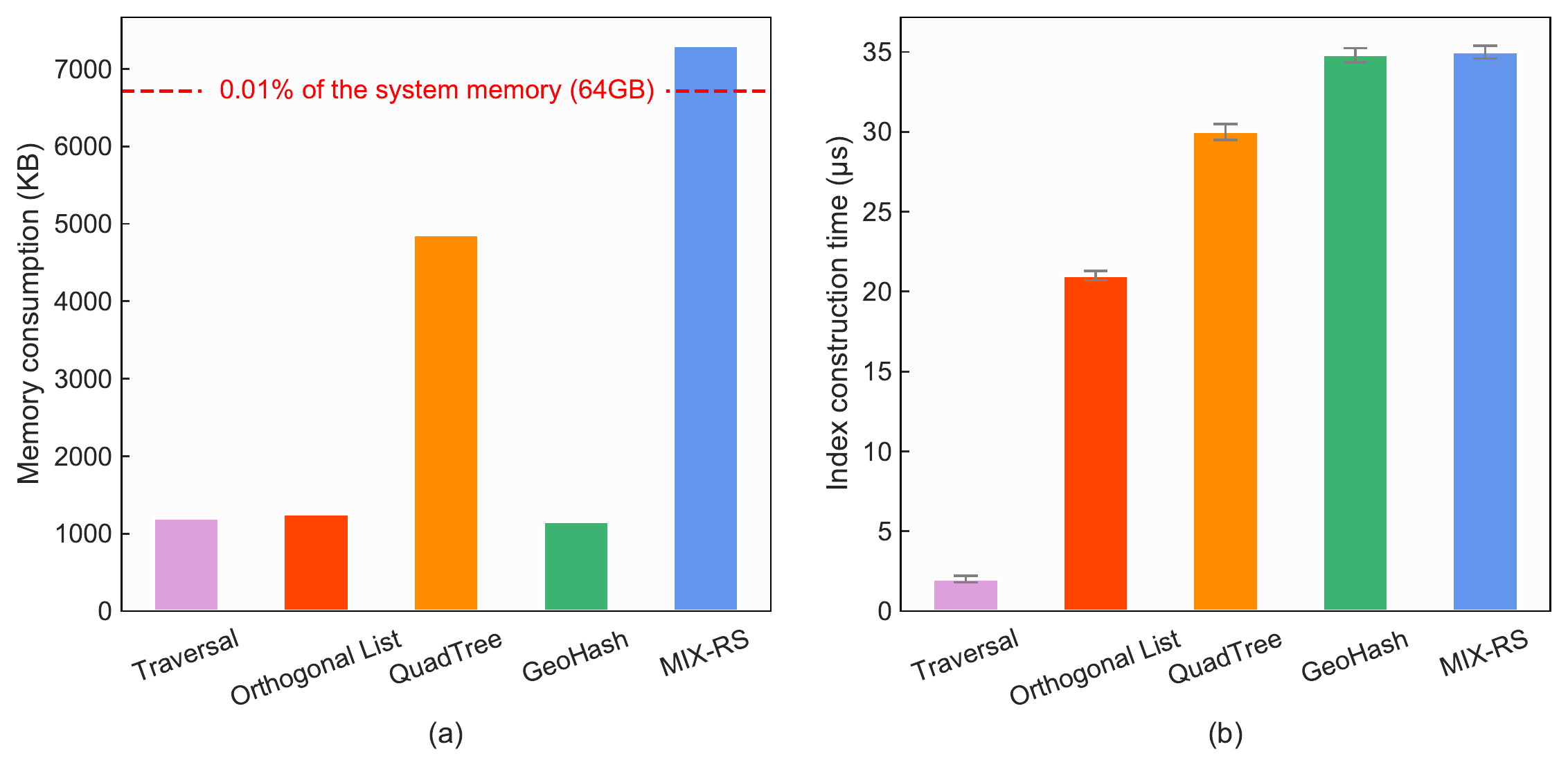}\\
    \caption{Evaluation of space and time overhead of multi-indexing and single indexing methods. (a) Amount of memory consumed by each method. The total system memory for each server is 64GB, and the red dotted line marks $0.01\%$ of the system memory. (b) Index construction time of each method, i.e., how long it takes for each method to complete the construction of index/indices it needs. }
    \label{fig:figure4}
  \end{center}
\end{figure*}

\subsection{Performance Evaluation of the Multi-indexing Mechanism}
\label{sec:sec4.2_performance_evaluation_of_multi_indexing_mechanism}

To demonstrate the excellent performance and efficiency of the proposed multi-indexing mechanism against single indexing methods, the experimental results of the comparison are shown in Fig. \ref{fig:figure3}. As illustrated in Fig. \ref{fig:figure3}(a), the line-chart clearly indicates that the multi-indexing mechanism outperforms all other single indexing methods when handling different number of random queries, indicated by the shortest amount of time elapsed. The GeoHash, Orthogonal List and QuadTree indexing methods cost nearly $2060\%$, $436\%$ and $60\%$ more processing elapsed time when processing different number of user queries than the multi-indexing mechanism, respectively. Therefore, the performance boost provided by the multi-indexing mechanism is very significant and can benefit the overall \emph{MIX-RS} framework in terms of RS data indexing. 

In terms of the scalability of the multi-indexing method, Fig. \ref{fig:figure3}(b) is the magnified version of Fig. \ref{fig:figure3}(a) without showing the brute force traversal method and hence the log-scale can be removed for better illustration. When the number of queries being processed keeps increasing, the total elapsed time increases linearly. The linear trend indicates that the method scales well in terms of query workloads. 

\subsection{Overhead of the Multi-indexing Mechanism}
\label{sec:sec4.3_overhead_assessment_of_multi_indexing_mechanism}

An excellent indexing design should enjoy superb performance and possess an acceptable overhead. To verify that the multi-indexing leveraged in the \emph{MIX-RS} does not have a severe overhead, the memory consumption and the index construction time are measured to demonstrate the method is space-efficient, and the index construction overhead is acceptable. 

As shown in Fig. \ref{fig:figure4}(a), the memory consumption of both the multi-indexing mechanism and single indexing methods are measured. It is natural to observe that the multi-indexing mechanism consumes most memory, which is around $7100$ KB. Apparently, constructing three indices should consume more memory than constructing a single index. Despite the highest memory consumption, the memory overhead of the multi-indexing mechanism is totally acceptable. Since for modern server infrastructures, nearly all servers are equipped with a memory that is orders of magnitude greater than the memory consumed by the multi-indexing mechanism. The servers we used have a memory of $64$ GB. As indicated by the red dotted line that marks $0.01\%$ of the system memory, the memory consumption of the multi-indexing mechanism only consumes approximately $0.011\%$ of the system memory. Therefore, the memory overhead is subtle and negligible. 

For the index construction time shown in Fig. \ref{fig:figure4}(b), the multi-indexing mechanism achieves nearly the same performance compared with the slowest single indexing method GeoHash, which is approximately $35$ microseconds. The three indices of the multi-indexing mechanism are constructed in parallel in each processing node, hence, it is natural to observe that the multi-indexing achieves nearly the same construction overhead with the slowest single method, which is not too long. Hence, the index construction overhead of the multi-indexing mechanism is not severe. 

Therefore, the multi-indexing mechanism brings significant performance improvement while not causing a noticeable overhead, demonstrating that using the multi-indexing mechanism is promising. 

\begin{figure*}[!t]
  \begin{center}
    \includegraphics[width=\textwidth]{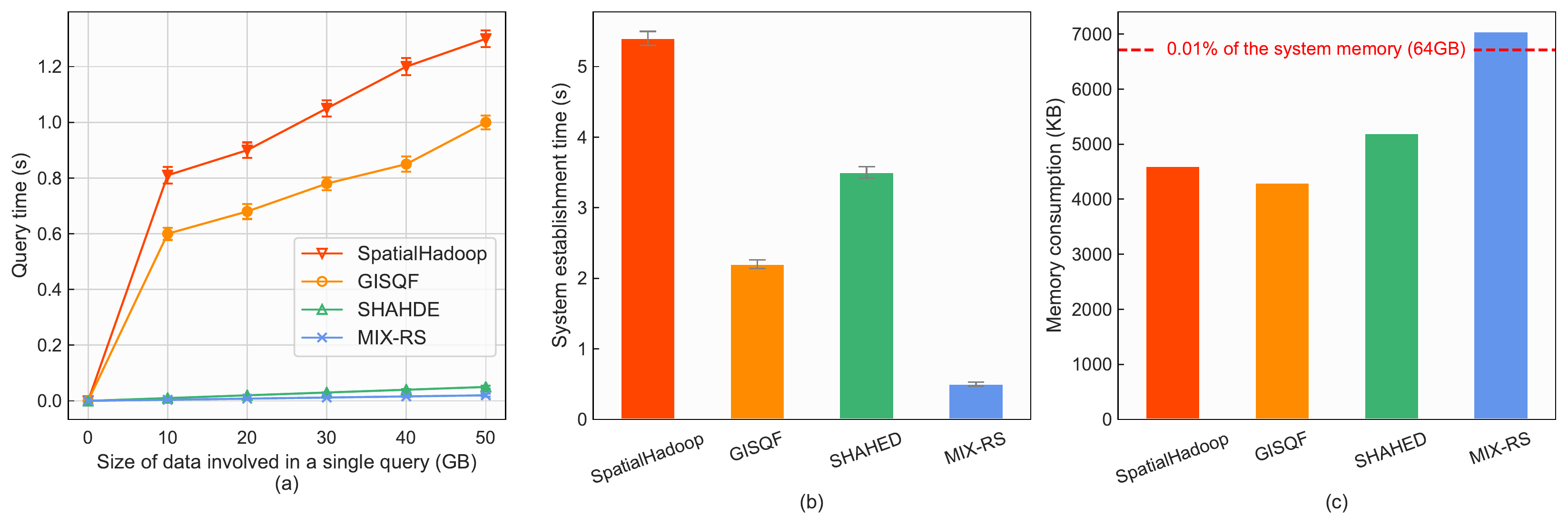}\\
    \caption{Performance and overhead comparisons at the framework level. (a) Amount of time taken to complete queries that involve different RS data sizes. The X-axis indicates the size of data involved in a single query, and the Y-axis is the query processing time. (b) System establishment time of each system to indicate the time overhead. (c) Memory consumption of each system to show the space overhead of each method. The total system memory for each server we used is $64$ GB, and the red dotted line marks $0.01\%$ of the system memory. }
    \label{fig:figure5}
  \end{center}
\end{figure*}

\subsection{Performance Evaluation at the Framework Level}
\label{sec:sec4.4_performance_evaluation_of_overall_system}

To verify the performance of the overall framework, we compare the \emph{MIX-RS} with three widely-used geospatial information systems, i.e., SpatialHadoop, GISQF and SHAHED. The performance comparison in terms of query processing time, system establishment time and system memory consumption are presented in Fig. \ref{fig:figure5}(a), (b) and (c), respectively. 

As shown in Fig. \ref{fig:figure5}(a), the \emph{MIX-RS} has the fastest query time among all frameworks. The SpatialHadoop, GISQF and SHAHED cost $64$, $49$ and $1.5$ times more query processing time when processing a relatively large query that involves around $50$ GB of RS data than the \emph{MIX-RS}, respectively. This is due to the \emph{MIX-RS} not only utilises multi-indexing on top of the HDFS with the benefit brought by data replication and parallel computation, but also avoids the warm-up overhead caused by the MapReduce, unlike other compared frameworks. Moreover, the query elapsed time linearly correlates with the size of the data involved in a single query, which justifies that the \emph{MIX-RS} presents excellent scalability when the amount of data involved in the query varies. 

As shown in Fig. \ref{fig:figure5}(b), the \emph{MIX-RS} is the fastest to be established, i.e., the time it takes from loading the data to finish constructing the index. The SpatialHadoop, GISQF and SHAHED cost $980\%$, $340\%$ and $600\%$ more system establishment time than the \emph{MIX-RS}. The evaluation result further demonstrates that the \emph{MIX-RS} stands out by having the lowest system establishment time overhead. 

In terms of system memory consumption overhead, it is natural to observe from Fig. \ref{fig:figure5}(c) that the \emph{MIX-RS} requires the highest amount of memory. However, compared with the memory capacity possessed by modern hardware infrastructures we used, the amount of memory consumed by \emph{MIX-RS} is only slightly greater than $0.01\%$ of the entire $64$ GB system memory, which is only a tiny portion. As such, despite naturally the \emph{MIX-RS} consumes the highest amount of memory, the ample space makes the overhead negligible. By demonstrating a low time and space overhead, the excellent performance and subtle overhead make the \emph{MIX-RS} very applicable.


\section{Conclusion}
\label{sec:sec5_conclusion}

In this paper, we propose the \emph{MIX-RS} framework for more efficient RS data indexing. The \emph{MIX-RS} naturally unifies the multi-indexing mechanism on top of the HDFS with data replication enabled. This unification is natural when data replication presents since it can utilise parallelism to boost the indexing performance. Hence, this holistic framework combines the fault tolerance provided by the HDFS and the indexing performance speedup provided by the multi-indexing mechanism. Moreover, the \emph{MIX-RS} framework requires very little effort in terms of framework implementation and has a low modification complexity. The framework causes very little time and space overhead to construct, making it feasible and applicable. Comprehensive experiments using real RS data are performed to verify the effectiveness of the multi-indexing mechanism, demonstrating the superior performance of the \emph{MIX-RS} system. Furthermore, both time and space overhead are evaluated, demonstrating the applicability of the \emph{MIX-RS} system.

\section*{Acknowledgement}
This work is supported in part by Key-Area Research and Development Program of Guangdong Province (2020B010164002) and the Fundamental Research Foundation of Shenzhen Technology and Innovation Council (Project No. KCXFZ20201221173613035).


\bibliographystyle{unsrtnat}

\bibliography{mix_rs}


\begin{strip}
\end{strip}

\begin{biography}[figure/bio-wu]
\noindent
\textbf{Jiashu Wu} received BSc. degree in Computer Science and Financial Mathematics \& Statistics from the University of Sydney, Australia (2018), and M.IT degree in Artificial Intelligence from the University of Melbourne, Australia (2020). He is currently pursuing his Ph.D at the University of Chinese Academy of Sciences (Shenzhen Institute of Advanced Technology, Chinese Academy of Sciences). His research interests include big data and cloud computing. 
\end{biography}

\begin{biography}[figure/bio-xiong]
\noindent
\textbf{JingPan Xiong} received his BSc. degree in software engineering from Wuhan Engineering University in 2017, and M.Sc degree in computer science from University of Chinese Academy of Sciences in 2020. He is currently working with Huawei Company, Shenzhen, China as computer engineer. His research interest is in big data storage, big data processing, and machine-learning applications. 
\end{biography}

\begin{biography}[figure/bio-dai]
\noindent
\textbf{Hao Dai} received the BS and M.Sc. degrees in Communication and Electronic Technology from the Wuhan University of Technology in 2015 and 2017, respectively. He is currently working toward the Ph.D. degree in the Shenzhen Institutes of Advanced Technology, Chinese Academy of Sciences. His research interests include mobile edge computing, federated learning and deep reinforcement learning. 
\end{biography}

\begin{biography}[figure/bio-wang]
\noindent
\textbf{Yang Wang} received the BSc degree in applied mathematics from Ocean University of China, in 1989, and the M.Sc. degree in computer science from Carleton University, in 2001, and the Ph.D degree in computer science from the University of Alberta, Canada, in 2008. He is currently in Shenzhen Institutes of Advanced Technology, Chinese Academy of Sciences, as a professor. His research interest includes cloud computing, big data analytics, and Java virtual machine on multicores. He is an Alberta Industry R\&D Associate (2009-2011), and a Canadian Fulbright Scholar (2014-2015). 
\end{biography}

\begin{biography}[figure/bio-xu]
\noindent
\textbf{Chengzhong Xu} received the Ph.D. degree from the University of Hong Kong in 1993. He is currently the Dean of Faculty of Science and Technology, University of Macau, China, and the Director of the Institute of Advanced Computing and Data Engineering, Shenzhen Institutes of Advanced Technology of Chinese Academy of Sciences.His research interest includes parallel and distributed systems and cloud computing. He has published more than 200 papers in journals and conferences. He serves on a number of journal editorial boards, including IEEE TC, IEEE TPDS, IEEE TCC, JPDC and China Science Information Sciences. He is a fellow of the IEEE.
\end{biography}

\end{document}